\documentclass[%
 aip,
 amsmath,amssymb,
 reprint,%
]{revtex4-1}

\usepackage{graphicx}% Include figure files
\usepackage{dcolumn}% Align table columns on decimal point
\usepackage{bm}% bold math
\usepackage{xcolor}
\usepackage{soul}
\usepackage[utf8]{inputenc}
\usepackage[T1]{fontenc}
\usepackage{mathptmx}
\usepackage{etoolbox}
\usepackage{hyperref}

\newcommand{\kater}[1]{\textcolor{black}{#1}}
\makeatletter
\def\@email#1#2{%
 \endgroup
 \patchcmd{\titleblock@produce}
  {\frontmatter@RRAPformat}
  {\frontmatter@RRAPformat{\produce@RRAP{*#1\href{mailto:#2}{#2}}}\frontmatter@RRAPformat}
  {}{}
}%
\makeatother

\begin{document}

\preprint{AIP/123-QED}

\title{Nitrogen Plasma Passivated Niobium Resonators for Superconducting Quantum Circuits}
% Force line breaks with \\

\author{K. Zheng}
\affiliation{Department of Physics, Washington University, Saint Louis, Missouri, 63130, USA}
\author{D. Kowsari}
\affiliation{Department of Physics, Washington University, Saint Louis, Missouri, 63130, USA}
\author{N. J. Thobaben}
\affiliation{Department of Physics, Saint Louis University, Saint Louis, Missouri, 63103, USA}
\author{X. Du}
\affiliation{Department of Physics, Washington University, Saint Louis, Missouri, 63130, USA}
\author{X. Song}
\affiliation{Department of Physics, Washington University, Saint Louis, Missouri, 63130, USA}
\author{S. Ran}
\affiliation{Department of Physics, Washington University, Saint Louis, Missouri, 63130, USA}
\author{E. A. Henriksen}
\affiliation{Department of Physics, Washington University, Saint Louis, Missouri, 63130, USA}
\affiliation{\mbox{Institute of Materials Science and Engineering, Washington University, Saint Louis, Missouri, 63130, USA}}
\author{D. S. Wisbey}
\affiliation{Department of Physics, Saint Louis University, Saint Louis, Missouri, 63103, USA}
\author{K. W. Murch}
\email[Email address: ]{murch@physics.wustl.edu}
\affiliation{Department of Physics, Washington University, Saint Louis, Missouri, 63130, USA}
\date{\today}% It is always \today, today,

\begin{abstract}
Microwave loss in niobium metallic structures used for superconducting quantum circuits is limited by a native  surface oxide layer formed over a timescale of minutes when exposed to an ambient environment. In this work, we show that nitrogen plasma treatment forms a niobium nitride layer at the metal-air interface which prevents such oxidation. X-ray photoelectron spectroscopy confirms the doping of nitrogen more than 5 nm into the surface and a suppressed oxygen presence. This passivation remains stable after aging for 15 days in an ambient environment. Cryogenic microwave characterization shows an average filling factor adjusted two-level-system loss tangent $\rm{F\delta_{TLS}}$ of $(2.9\pm0.5)\cdot10^{-7}$ for resonators with 3 $\rm{\mu}$m center strip and $(1.0\pm0.3)\cdot10^{-7}$ for 20 $\rm{\mu}$m center strip, exceeding the performance of unpassivated samples by a factor of four.
\end{abstract}

\maketitle

\begin{figure}
\centering
\includegraphics{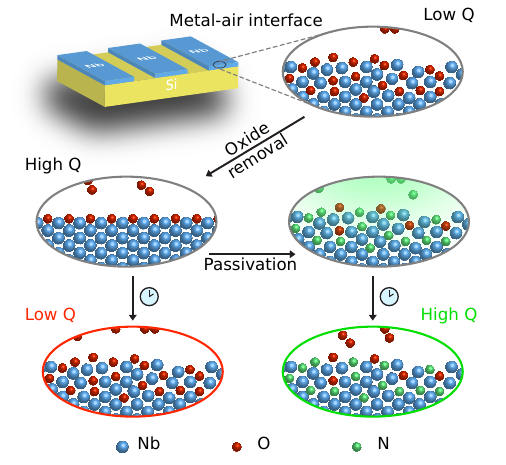}
\caption{{\bf Nitrogen passivation of Nb resonators.}   Nb films grow an amorphous oxide layer at the MA interface that results in low quality factor. Removal of this oxide with a BOE clean improves Q, but the oxide regrows over time (left).  Passivating the MA interface with N inhibits the growth of oxides, maintaining high Q over time.} \label{fig:flow}
\end{figure}

Increasing the coherence time of superconducting qubits while maintaining reasonable gate speeds enables more powerful quantum processors\cite{supremacy,IBM64}. The improvement of fabrication techniques plays an important role in this effort\cite{OliverMRS,materialMatters}. The fabrication process of high coherence planar superconducting quantum processors now typically involves two superconducting layers. The first layer makes up the ground plane and all the circuit elements other than the Josephson junction\cite{Josephson,RowellJJ} (JJ) while the second layer defines the JJs. The two-layer process allows optimization of the quality factor of the superconducting capacitor pads  independent from the requirements of the double-angle-evaporated aluminum JJ layer. Microwave coplanar waveguide (CPW) resonators have been demonstrated as a robust platform to characterize the microwave loss in superconducting materials\cite{MartinisDielectric, McRaeReview}, and have been instrumental in the continuous improvement of film and device quality\cite{rigettiNb, HouckTa}. \kater{In typical device geometries, less than 1 percent of the electric field energy is stored in the thin dielectric layers at the metal-air (MA), metal-substrate (MS), and substrate air (SA) interfaces}\cite{Wenner, oliverAlTiN}.%\kater{Usually the majority of the electric field energy of a superconducting device is stored in the substrate and the vacuum with less than 1 percent in the thin dielectric layers at the metal-air (MA), metal-substrate (MS), and substrate air (SA) interfaces.}
 \kater{  These dielectric layers may host two-level-system (TLS)}\cite{MullerTLS} \kater{defects resulting in a high loss tangent} \cite{MartinisDielectric, Wenner, oliverAlTiN} \kater{that dominates the single-photon energy loss of state-of-the-art superconducting devices, despite the small participation of these dielectric layers.} For Nb resonators, \kater{the primary TLS loss comes from the oxide layer}\cite{NbODFT,NbSurfaceEPR} \kater{at the MA interface. As the electric field energy stored in a thin dielectric layer at the interface is proportional to its thickness}\cite{VerjauwNb}\kater{, the removal of the oxide %reduces the energy that is stored in the lossy oxide layer and 
 results in lower microwave loss. The removal of the surface oxide layer in Nb CPW resonators has resulted in}  single-photon internal quality factors $\rm{Q_i}$ up to 5 million\cite{SiddiquiNb} and filling factor adjusted two-level-system loss tangents $\rm{F\delta_{TLS}}$ down to $9\cdot10^{-8}$. However, the oxide grows back following a Cabrera-Mott\cite{CabreraMott} behavior within several hours, reintroducing TLSs at the MA interface\cite{VerjauwNb, NbXRR} (Fig.~\ref{fig:flow}). As the buffered oxide etch (BOE) used to etch $\rm{NbO_x}$ also etches the Al JJ\cite{etchRate}, it is difficult to incorporate Nb resonators with a low density of TLSs at the MA interface into superconducting quantum circuits. 

Nitride superconductors such as NbN and TiN are known to only have a thin native surface oxide layer at room temperature and therefore are expected to have fewer TLSs at the MA interface\cite{NbTiNOxide}. \kater{As for the MS interface, recent studies}\cite{wisbeyTiN, SUNY_fab, ALDTiN, OliverTSV} \kater{have shown that high quality TiN (200) films can be deposited directly on Si substrate without a $\rm{SiN_{x}}$ seed layer deposited in a separate process step which reduces the total microwave loss. However, the microwave loss of TiN at the MS interface is still higher than elemental superconductor and work is needed to further improve the quality of the MS interface}\cite{oliverAlTiN}. \kater{Resonators with $\rm{Q_i}$ on the order of 1 million have been fabricated using sputtered and atomic-layer-deposited NbN thin films}\cite{NbNRes, ALDNbN}\kater{, and NbN-based superconducting qubits have reached a relaxation time of 16 $\rm{\mu s}$}\cite{NbNQubit}. \kater{However, the performance of these devices is still below the state of the art necessitating further investigation of NbN films.} If a resonator could have the MA interface from a nitride superconductor and the MS interface from an elemental superconductor, then its microwave loss caused by TLS is expected to be extremely low. There have been attempts to deposit a thin TiN film covering Nb to prevent the growth of $\rm{NbO_x}$\cite{TiNonNb}, but the sidewalls are not covered by TiN and can still oxidize.

Nitrogen doping\cite{N_dope} is a well-established technique used to create low-loss Nb 3D cavities. By baking the Nb cavity in $\rm{N_2}$ gas at a temperature above 800 $\rm{^\circ}$C, a conformal layer of $\rm{NbN_x}$ is formed at the MA interface which inhibits the growth of surface oxide\cite{SRFNb_review}. In theory, this method can be adapted to make planar superconducting devices, however the effect of high temperature is not yet well understood. Treatment at high temperature can change the stress and grain size of the deposited metal and allow Nb to diffuse into the Si substrate, which all could potentially affect microwave loss\cite{MartinisAl, HouckNb,CameronDLTS}. Baking at high temperature can also introduce dislocation and vacancy defects in the Si substrate, whose effects on microwave loss have not yet been systematically explored.

In this Letter, we demonstrate nitrogen doping of Nb CPW resonators without introducing potentially performance-limiting defects related to high temperature. We utilize a radio frequency (RF) plasma to satisfy the activation energy required for nitrogen doping. We show that a $\rm{N_2}$ plasma at 300 $\rm{^\circ}$C for 10 minutes dopes the top 5 nm of the Nb surface with N which suppresses the growth of Nb oxides at the surface. Furthermore, we demonstrate that the passivation remains robust in an ambient air environment for sufficiently long periods of time to incorporate passivated Nb structures into complex, multi-layer quantum processors. Our process can be easily manifested with equipment available in typical industrial and academic facilities.

The films used in this study are fabricated by first cleaning a 2-inch instrinsic Si (111) wafer\footnote{Since Nb does not match to the lattice of Si (111) or  Si (100), we expect similar results independent of wafer orientation.}  with resistivity greater than 8000 $\rm{\Omega\cdot}$cm in piranha solution (3:1 mixture of $\rm{H_2SO_4}$ and $\rm{H_2O_2}$) at 120 $\rm{^\circ}$C for 10 minutes followed by a BOE clean (7:1 mixture of $\rm{NH_4F}$ and HF)  for 5 minutes. We then load the wafer into an electron-beam evaporation chamber (AJA ATC-ORION-8E) in less than 3 minutes.  We evaporate 250 nm of Nb ($99.95\%$ purity) at  a  chamber pressure of $\rm{3\cdot10^{-9}}$ Torr and a rate of 2 nm/min\cite{ebeamNb, OurNb} at room temperature, \kater{resulting in a thickness uniformity better than 2.5\% variation. The thickness of the Nb film is confirmed by crystal microbalance and profilometer measurements.}

The resonator pattern is defined after we deposit the Nb film and before the passivation steps. We utilize a positive photoresist (MicroChem Shipley S1805) and a 375 nm optical direct-write laser lithography system (Heidelberg DWL 66+) to pattern the film. We etch the patterned film  using a mixture of \text{$\rm{SF_6}$} and Ar in an RIE system (Oxford Plasma Lab 100 ICP). Our etching recipe produces a \kater{$\rm{550\pm20}$} nm deep trench in the Si substrate. We then oxygen plasma ash the samples (100 W, 15 sccm $\rm{O_2}$) for 30 s and finally soak them in MicroChem Remover PG for $>8$ \kater{hours} at 70 $\rm{^{\circ}C}$ to remove the remaining resist.
\begin{figure}
\centering
\includegraphics{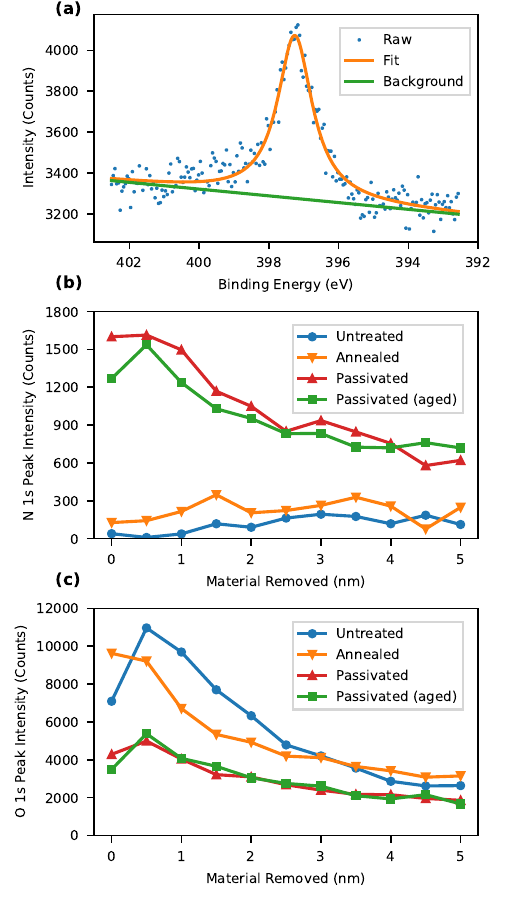}% Here is how to import EPS art
\caption{{\bf \kater{Effect of different surface treatments on the XPS 1s peak of N and O}} (a) The observed N 1s peak for a passivated Nb film at 1 nm depth. (b, c) The N 1s peak (b) and the O 1s peak (c) for different film treatments as a function of etch depth by in-situ Ar ion milling.} \label{fig:XPS}
\end{figure}

\begin{figure*}
\centering
\includegraphics{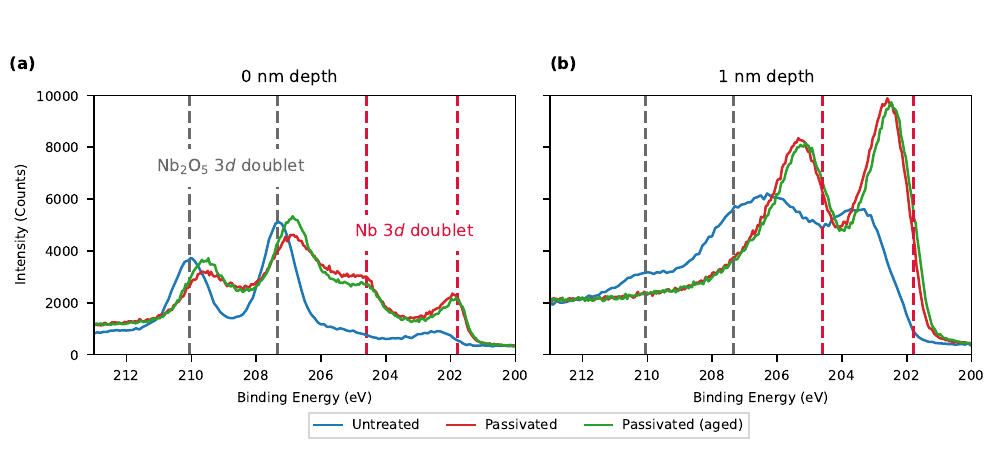}
\caption{XPS spectrum of the Nb 3d peaks with (a) 0 nm, and (b) 1 nm of surface material removed. } \label{fig:XPS2}
\end{figure*}
To passivate the samples, we start by performing a BOE clean for 30 minutes to remove all the surface oxide before transferring the wafer into a plasma enhanced chemical vapor deposition (PECVD) chamber (Oxford Instruments Nanofab). The transfer time between removal from the BOE solution and evacuation of the PECVD chamber is kept below  3 minutes. We first let the wafer sit at 300 $\rm{^\circ}$C for 10 minutes with 1200 sccm $\rm{N_2}$ flow and a chamber pressure of 1 Torr. We then start a 20 W RF plasma at the same chamber condition for 10 minutes. Lastly, we cool down the chamber to 100 $^\circ$C with 100 sccm of $\rm{N_2}$ flow over the course of one hour before removing the sample. \kater{The passivation process forms a $\rm{\sim 2}$~nm thick $\rm{SiN_x}$ layer at the exposed Si surface in the trench area.} We perform a third BOE clean for 15 s to remove \kater{this} thin $\rm{SiN_{x}}$ layer. 

The PECVD system uses a routine process to remove oxides from the chamber walls. This process introduces residual F in the chamber which reacts with the Nb films. To mitigate contamination from F, we deposit $\sim 1\ \mu$m of $\rm{SiO_x}$ on a dummy wafer and the chamber wall immediately before passivating our sample.

We characterize the surface atomic species of passivated and unpassivated Nb films using X-ray Photoelectron Spectroscopy (XPS). The XPS system (Physical Electronics 5000 VersaProbe II Scanning ESCA Microprobe) uses an Al $\rm{K_{\alpha}}$ line source and  in-situ Ar ion milling to facilitate analysis of material at depth. Figure \ref{fig:XPS}(a) shows the XPS scan of a passivated Nb film with its top 1 nm removed. The observed peak corresponds to N 1s electrons. We fit the N 1s peak to a psuedo-Voigt function and the background to a linear function. We use the same practice to fit all the N 1s and O 1s\cite{doublePeakO} peaks measured in this study.

Figure \ref{fig:XPS}(b) shows the area under the N 1s peak as a function of surface material removed for Nb films with different treatments. The untreated sample and the sample that is simply annealed in $\rm{N_2}$ at 300 $\rm{^\circ C}$ without lighting a plasma do not show a strong N presence. In comparison, plasma passivated Nb film shows a well-defined N peak that persists to a depth of 5 nm, which suggests successful incorporation of N atoms. The intensity of the N peak does not change significantly after we age the passivated Nb sample in an ambient environment for 15 days. In Fig.~\ref{fig:XPS}(c), we show that the presence of O in the surface 5 nm is reduced for passivated samples. Similar to N, the O concentration in the plasma passivated sample does not change after being aged in an ambient environment for 15 days.

The study of N and O 1s electrons confirms a suppressed oxygen concentration near the surface for passivated samples.  We now examine the Nb 3d electrons to investigate the oxides present near the surface of the films.  Figure \ref{fig:XPS2} shows the raw XPS spectrum of the different chemical states of the Nb 3d peaks from the same samples shown in Fig.~\ref{fig:XPS}. As the stoichiometry of the oxide layer is complex\cite{onOxNb} and we have also introduced a nitride with unknown stoichiometry, we expect the spectrum to contain at least 10 individual peaks that overlap with each other. \kater{We examine the trend of the energy shift of the dominant peaks from an oxygen-rich high binding energy to lower binding energies which are associated with states that are less oxygen-rich.} At zero depth [Fig.~\ref{fig:XPS2}(a)] we observe clear Nb 3d doublet peaks corresponding to $\rm{Nb_2O_5}$ \kater{at 207.5 and 210 eV} for the unpassivated \kater{sample}. \kater{These two peaks shift to lower energies of 206.8 and 209.3 eV for the passivated films which correspond to a less oxygen-rich Nb oxide state. The un-oxidized Nb 3d doublet peaks are also clearly present at 201.8 and 204.3 eV for the passivated sample.} As shown in Fig.~\ref{fig:XPS2}(b), \kater{the unpassivated sample has two dominate peaks at $\sim$203.5 and $\sim$206.5 eV and a small peak at 210 eV. We associate this to a combination of $\rm{NbO}$, $\rm{Nb_2O_5}$, and a large population of $\rm{NbO_2}$. In comparison, the dominant peaks of the passivated sample are at 202.6 and 205.3 eV, which we attribute to a majority of NbO and NbN together with a small amount of Nb.} %The Nb spectrum at both 0 and 1 nm depths confirm our observation of suppressed O presence in the passivated sample from the O 1s peaks. } 
We also observe that \kater{the Nb XPS spectrum does not change significantly for the passivated sample after 15 days of aging in an ambient environment. This indicates that the passivated sample does not accumulate significantly more oxide over this period of time.}

\begin{figure}
\centering
\includegraphics{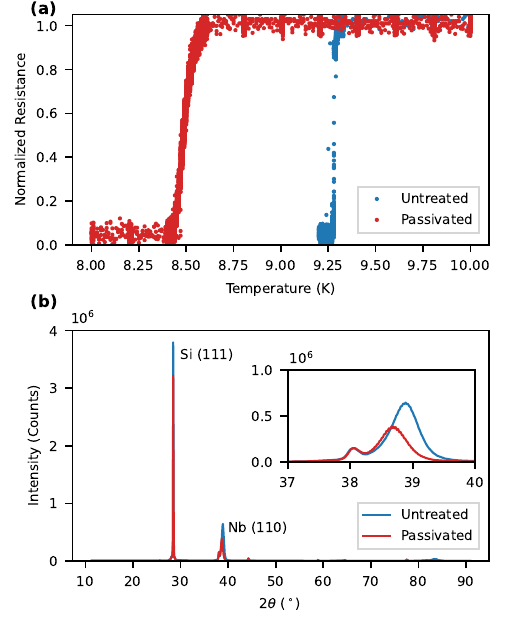}
\caption{(a) Superconducting transition of plasma passivated and untreated samples. (b) XRD $\rm{2\theta}$ scan of plasma passivated and untreated Nb films. The inset shows the Nb (110) peaks from Cu $\rm{K_{\alpha1}}$ and $\rm{K_{\alpha2}}$.}
\label{fig:Tc}
\end{figure}

To study the effect of the passivation process on the quality of the Nb films, we measure the direct-current (DC) resistance of the passivated and the untreated films as a function of temperature using a 4-point technique inside a Quantum Design physical property measurement system (PPMS). Both samples are diced from the same Nb film deposited on the same Si (111) wafer allowing us to attribute the change in film properties to the passivation process alone. We extract the residual resistivity ratio (RRR) by taking the ratio between the resistance value at 297 K and 10 K. We find that the untreated sample has a RRR of 4.86, which is similar to our previous result on Si (100)\cite{OurNb}. The plasma passivated sample has a reduced RRR value of 2.96.  Figure~\ref{fig:Tc}(a) shows the resistance near the superconducting transition of both samples. We observe a superconducting critical temperature $\rm{T_c}$ of $\rm{9.28\pm0.05}$ K for the untreated sample, and a suppressed $\rm{T_c}$ of $\rm{8.49\pm0.05}$ K for the plasma passivated sample with $\rm{T_c}$ taken from the temperature value when the resistance drops to half of its residual value in the normal state. \kater{The suppressed  $\rm{T_c}$ can be explained by assuming that the top $\rm{NbN_x}$ passivation layer is disordered and therefore has a much lower $\rm{T_c}$ than the Nb film. Given the bilayer system we can estimate the thickness of $\rm{NbN_x}$ to be about 14.5 nm in agreement with our observation from the XPS data that N atoms extend at least 5 nm into the surface.  }\cite{Cooper, NbPenetration}.  \kater{We further estimate the zero temperature penetration depth to be 77 nm for the untreated film and 97 nm for the passivated film} \cite{tinkham, NbPenetration,vanDuzer}.
%\begin{equation}
%\mathcolorbox{yellow}{\mathrm{T_c=T_{c0}e^{\frac{\lambda-\Delta d}{NV(d-\Delta d)}}}},
%\end{equation}
%\kater{where $\rm{T_{c0}}$ is the critical temperature of the superconductor, d the thickness of the superconducting film, $\rm{\Delta d}$ the thickness of the disordered film, NV is the bulk interaction potential, and $\rm{\lambda}$ the mean free path of the superconductor. Using NV of 0.32 for Nb}\cite{NbPenetration}\kater{, $\rm{\lambda}$ of 8 nm}\cite{NbRRR} \kater{estimated from the RRR of the passivated sample, and $\rm{T_{c0}}$ of 9.28 K from our unpassivated film, we estimate $\rm{\Delta d}$ to be 14.5 nm which agrees with our observation that N atoms extend at least 5 nm into the surface from the XPS data. We estimate the zero temperature penetration depth}\cite{tinkham, NbPenetration}\kater{ to be 77 nm for the untreated film and 97 nm for the passivated film based on the coherence length calculated from $\rm{T_c}$, $\rm{\lambda}$ calculated from RRR, and a zero temperature London penetration depth of 38 nm for Nb}\cite{vanDuzer}. 

Figure~\ref{fig:Tc}(b) shows the X-ray diffraction (XRD) spectrum of both samples measured in a Rigaku MiniFlex system with a Cu $\rm{K_\alpha}$ source. Although both samples show similar texture, the Nb (110) peak of the passivated sample has about half of the intensity compared to the untreated film. \kater{We do not observe any peaks that are associated to $\rm{NbN_x}$, which confirms our assumption that the $\rm{NbN_x}$ layer is disordered.}

We utilize cryogenic microwave measurements to extract the power-dependent quality factor of CPW resonator structures, and to study the effect of passivation on the TLS-induced loss in the devices. We apply the passivation process to the  resonators after the CPW structures are defined with RIE so that the sidewalls of the resonators are also passivated. The resonators are capacitively coupled to a transmission line with coupling quality factor $\rm{Q_c}\simeq 0.5\times 10^6$. Hanger-style resonators\cite{WisbeyB4C, OurNb} with two types of geometries are fabricated. The first geometry has a center strip width of 3 $\rm{\mu}m$, gap width of 2 $\rm{\mu}m$, and has a continuous ground plane. This design allows us to clearly observe the effect of TLSs by maximizing capacitive dielectric loss\cite{MullerTLS} and minimizing inductive\cite{MartinisVortex, resQP} and radiative\cite{Huang_PRX} losses. The second geometry has center strip width of 20 $\rm{\mu}$m, gap width of 12 $\rm{\mu}$m, and has holes (5$\times$5 $\mu$m$^2$ squares) periodically (every 15 $\mu$m) etched into the ground plane\cite{PlourdeTrap}. This design minimizes the capacitive loss but is more susceptible to radiative loss\cite{Huang_PRX} and loss caused by trapped magnetic flux\cite{MartinisVortex}. We use GE varnish to attach each resonator sample to a microstrip-style microwave launch with Pd-coated Cu metallization and Rogers Duroid 6010LM dielectric.  We make 3 wirebonds to each port with 1 mil Al-$\rm{1\%}$Si bonding wire. The microstrip launch is enclosed inside a Pd-coated Cu sample package. The package is cooled inside a Rainier Model 103 adiabatic demagnetization refrigerator (ADR) down to 50 mK. We add $\rm{\sim}$100 dB of attenuation distributed among different temperature stages to the input line\cite{WisbeyB4C}. We shield the sample from stray magnetic fields with a high permeability can (Amuneal Cryoperm) surrounding the sample package. The output of the sample passes through one circulator to a high electron mobility transistor (HEMT) amplifier at the 4 K stage. Further amplification is used at room temperature. 

A vector network analyzer (VNA) is used to measure the transmission through the sample. The $\rm{Q_i}$ of each resonator is extracted at different average circulating photon numbers $\rm{\langle n \rangle}$ ranging from \textasciitilde1 to \textasciitilde$\rm{10^7}$ using the diameter correction method\cite{DCM} (DCM). $\rm{F\delta_{TLS}}$ and the high-power internal quality factor $\rm{Q_i^{HP}}$ are calculated from\cite{McRaeReview}
\begin{equation}
\mathrm{F\delta_{TLS}}\frac{\mathrm{tanh}(\frac{\rm{hf}}{\rm{k_BT}})}{\sqrt{1+\frac{\mathrm{\langle n \rangle}}{\mathrm{n_c}}}}=\frac{1}{\mathrm{Q}_\mathrm{i}^\mathrm{HP}} - \frac{1}{\mathrm{Q}_\mathrm{i}},
\end{equation}
where $\rm{h}$ is the Planck constant, $\rm{f}$ is the resonant frequency of the resonator, $\rm{k_B}$ is the Boltzmann constant, $\rm{T}$ is the temperature of the resonator, and $\rm{n_c}$ is the critical photon number that differentiates the high and low power regions. When $\rm{\langle n \rangle}$ is large, the contribution to microwave loss from saturable TLSs approaches zero, so $\rm{Q_\mathrm{i}^\mathrm{HP}}$ gives a good estimate of the other sources of loss which mainly result from quasiparticles, trapped magnetic flux, and radiation.  

\begin{figure}
\centering
\includegraphics{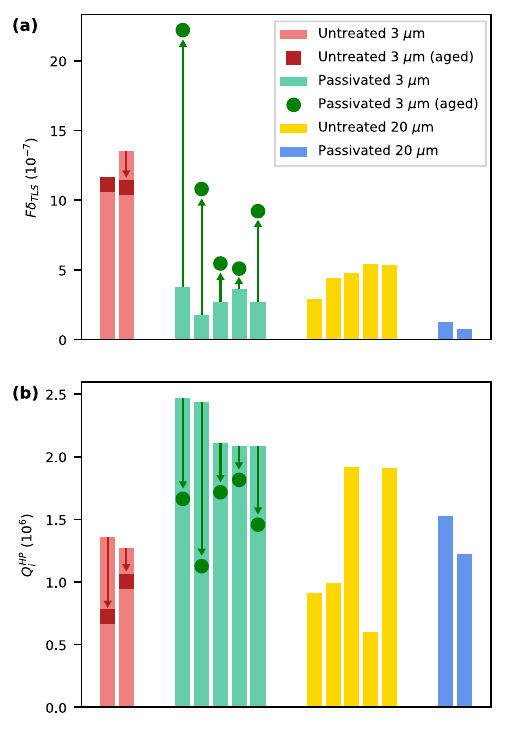}% Here is how to import EPS art
\caption{{\bf Microwave performace of passivated and untreated samples.} (a) Extracted $\rm{F\delta_{TLS}}$ values among plasma passivated and untreated resonators with 3 $\mu$m and 20 $\mu$m center strip geometries. (b) Extracted $\rm{Q_i^{HP}}$ of the same resonators shown in (a).} \label{fig:qvalues}
\end{figure}

Figure \ref{fig:qvalues}(a) shows a comparison of the $\rm{F\delta_{TLS}}$ value among plasma passivated and untreated samples of both designs. Untreated resonators with the 3 $\mu$m center strip geometry have an average $\rm{F\delta_{TLS}}$ of $(12.5\pm1.5)\cdot10^{-7}$. This agrees with the result for untreated samples obtained in our previous study\cite{OurNb}, which uses the same design but intrinsic Si (100) wafers instead of Si (111). In comparison, the passivated resonators have an average $\rm{F\delta_{TLS}}$ of $(2.9\pm0.5)\cdot10^{-7}$, which is $\rm{\sim}$4 times lower than the untreated value. Resonators with the 20 $\mu$m center strip geometry have an average $\rm{F\delta_{TLS}}$ of $(4.6\pm1.0)\cdot10^{-7}$ for untreated samples and $\rm{F\delta_{TLS}}$ of $(1.0\pm0.3)\cdot10^{-7}$ for passivated samples. Figure \ref{fig:qvalues}(b) shows the extracted $\rm{Q_i^{HP}}$ of the same resonators shown in Fig.~\ref{fig:qvalues}(a). Resonators with the 3 $\mu$m center strip geometry show an increased $\rm{Q_i^{HP}}$ compared to untreated devices. We attribute this \kater{potentially} to a reduced TLS density in the regions of the sample that are too far from the resonators to be effectively saturated at high power \kater{and a reduced spin defects residing in the $\rm{Nb_2O_5}$ layer}\cite{NbODFT}. The $\rm{Q_i^{HP}}$ values of the resonators with the 20 $\mu$m center strip geometry are less uniform. We suspect that this could be due to their susceptibility to trapped vortices and radiation to lossy regions inside the package. The $\rm{Q_i^{HP}}$ values of our resonators are lower compared to other similar studies\cite{SiddiquiNb, MartinisAl}, which we attribute to specific details of the sample package\cite{Huang_PRX, Pdres} and measurement setup. %the  Pd coating and GE varnish inside our sample package. Although Pd coating protects Cu from oxidation, its resistance is much higher than Cu\cite{Pdres} which leads to higher microwave loss\cite{Huang_PRX}. 

%and $Q_\mathrm{i}^\mathrm{HP}$

To test the stability of the passivated samples over time, we allow some of the samples to age in an ambient environment for 45 days. Figure \ref{fig:qvalues}(a) indicates the change in  $\rm{F\delta_{TLS}}$ after this aging duration.  We observe negligible change in $\rm{F\delta_{TLS}}$ for the untreated devices, indicating that the oxide had already reached self-limited thickness before the first measurement. The passivated resonators exhibit increased $\rm{F\delta_{TLS}}$ of varying degrees. In particular, two of the five passivated resonators maintained a significantly lower $\rm{F\delta_{TLS}}$ compared to the untreated resonators, validating the stability of the passivated samples. Three of the passivated resonators, however, exhibited a significant increase in $\rm{F\delta_{TLS}}$, which could be due to additional contamination of the resonators during storage. Visual inspection reveals significant residue and debris in the gap regions of CPWs for the resonators with significantly increased $\rm{F\delta_{TLS}}$.  \kater{We attribute this debris to the process of removing, storing, and re-gluing samples into the sample package for the second measurement.} %\kater{Sonicating the samples in acetone does not remove the residue. We did not attempt cleaning steps that are more agressive which could introduce additional sources of microwave loss.}
In  Fig. \ref{fig:qvalues}(b) we indicate the change in $\rm{Q_\mathrm{i}^\mathrm{HP}}$ for the aged devices. For all samples we observe a reduction in $\rm{Q_\mathrm{i}^\mathrm{HP}}$ after aging. % the r   of  do not show a significant change in their $\rm{F\delta_{TLS}}$ values. For the passivated sample, we observe that 2 out of the 5 resonators measured after 45 days still retain a $\rm{F\delta_{TLS}}$ value much lower than the average untreated value. We observe that the other 3 resonators have a dramatic increase in their $\rm{F\delta_{TLS}}$. After removing the sample after the aging study, we inspect the resonators under an optical microscope. We find that there are organic-looking residuals in the trenches of the resonators whose $\rm{F\delta_{TLS}}$ value increases dramatically. We are not able to find similar residuals on our untreated chip. Figure 5(b) shows an example of the residuals we see. We are not able to effectively remove the residual by sonicating the sample inside acetone and isopropyl alcohol. 

%\begin{figure}
%centering
%\includegraphics{Fig5.pdf}% Here is how to import EPS art
%\caption{(a) Aging study of the resonators with 3 (2) geometry. (b) Optical micro-graph of the dirty trench region of a passivated resonator whose $\rm{F\delta_{TLS}}$ increases dramatically in the aging study.} \label{fig:aging}
%\end{figure}
\kater{We extract the filling factor of the MA and MS interface based on a simulation of the cross section of the resonators with 3 $\rm{\mu m}$ center strip using HFSS Q3D extractor. By assuming a dielectric constant (thickness) of 41 (5 nm) for the MA interface and of 4 (2 nm) for the SA interface we obtain a filling factor of $\rm{2.3\cdot10^{-5}}$ for the MA interface and $\rm{5.0\cdot10^{-4}}$ for the SA interface, both of which scale linearly with their thickness. By assuming a loss tangent of $\rm{9.9\cdot10^{-3}}$ for $\rm{Nb_2O_5}$ and $\rm{1.7\cdot10^{-3}}$ for $\rm{SiO_x}$}\cite{VerjauwNb, oliverAlTiN}\kater{, we obtain a total loss of $\rm{1.08\cdot10^{-6}}$, which agrees with our results of the untreated devices. The microwave loss of the passivated devices agrees with the loss estimated with a 1.3 nm thick $\rm{Nb_2O_5}$ layer and a 0.5 nm thick $\rm{SiO_x}$ layer. As we do not observe significant change in the O population from the XPS scan, we assume the increase of the microwave loss comes mainly from the increase of $\rm{SiO_x}$ thickness from 0.5 nm to 1 nm, which is expected from exposing Si to ambient conditions}\cite{SiOx}\kater{.} 

Our study demonstrates a recipe for passivation of Nb structures with nitrogen plasma which dopes the top 5 nm of the Nb surface with N atoms. These N atoms suppress the O concentration and therefore reduce the amount of TLS defects that induce microwave loss. The N and O populations are stable after 15 days of aging in ambient air according to XPS measurements. Cryogenic microwave measurements confirm that our passivation process reduces the microwave loss of CPW resonators of two different design geometries. Our process removes the stringent time sensitivity associated with the regrowth of $\rm{NbO_x}$ and allows for the incorporation of low microwave loss Nb structures into state-of-the-art quantum processors. Our process also creates a platform on which other sources of microwave loss can be further studied. Although $\rm{T_c}$ and XRD measurements suggest that our process creates a more disordered superconductor, which may increase the susceptibility to losses caused by quasiparticle and magnetic vortices, these mechanisms do not yet dominate the microwave loss. While the side effect of suppressed superconductivity may be reduced by starting with Nb films with large grain size \kater{and longer electron mean free path}, the suppressed superconductivity may also be useful in producing low loss kinetic-inductance-based sensors\cite{ZmuidzinasReview, WisbeyKid}. As Ta also has a superconducting nitride\cite{TaN} and it is more refractory than Nb, our passivation process may also work in reducing the microwave loss of Ta.

\begin{acknowledgments}
This research is supported by NSF Grant No. PHY-1752844 (CAREER), AFOSR MURI Grant No. FA9550-21-1-0202, DOE grant DE-SC0017987, and the John Templeton Foundation, Grant No. 61835. The authors acknowledge the use of facilities at the Institute of Materials Science and Engineering in Washington University.

\end{acknowledgments}

\section*{Data Availability Statement}
The data that support the findings of this study are available from the corresponding author upon reasonable request.

\appendix


%\nocite{*}
%\bibliography{ref}% Produces the bibliography via BibTeX.

\end{document}